\documentclass[11pt]{article}

\usepackage[a4paper,margin=2.5cm]{geometry}
\usepackage[auth-sc-lg,affil-sl]{authblk}
\usepackage[algoruled,vlined]{algorithm2e}
\usepackage{nomencl}
\usepackage{amsmath}
\usepackage{xcolor}
\usepackage{graphicx}
\usepackage{hyperref}

\newcommand{\rev}[1]{#1}

\newcommand{\Fref}[1]{Figure~\ref{#1}}
\newcommand{\Eref}[1]{Eq.~\eqref{#1}}
\newcommand{\Aref}[1]{Appendix~\ref{#1}}

\newcommand{\M}[1]{{\boldsymbol #1}}

\newcommand{\lay}[1]{^{(#1)}}
\newcommand{\rot}[1]{\varphi\lay{#1}}
\newcommand{\rote}[2]{\varphi\lay{#1}_{#2}}
\newcommand{\Drot}[2]{\Delta\rot{#1}_#2}

\newcommand{\x}[1]{x\lay{#1}}
\newcommand{\X}[1]{X\lay{#1}}
\newcommand{\z}[1]{z\lay{#1}}
\newcommand{\Z}[1]{Z\lay{#1}}
\newcommand{\F}[2]{F\lay{#1}_{#2}}
\newcommand{\U}[2]{U\lay{#1}_{#2}}
\newcommand{\Ht}[2]{H\lay{#1}_{#2}}

\newcommand{\h}[1]{h\lay{#1}}

\newcommand{\uc}[1]{u_0\lay{#1}}
\newcommand{\uce}[2]{u_{0,#2}\lay{#1}}
\newcommand{\Duc}[2]{\Delta u_{0,#2}\lay{#1}}
\newcommand{\wc}[1]{w_0\lay{#1}}
\newcommand{\wce}[2]{w_{0,#2}\lay{#1}}
\newcommand{\Dwc}[2]{\Delta w_{0,#2}\lay{#1}}
\newcommand{\Sc}[1]{E\lay{#1}}
\newcommand{\Kc}[1]{K\lay{#1}}
\newcommand{\Gc}[1]{\Gamma\lay{#1}}

\newcommand{\MF}[1]{\M{F}\lay{#1}} % Deformation gradient
\newcommand{\MR}[1]{\M{R}\lay{#1}} % Rotation tensor
\newcommand{\MU}[1]{\M{U}\lay{#1}} % Stretch tensor
\newcommand{\MH}[1]{\M{H}\lay{#1}}

\newcommand{\half}{\mbox{$\frac{1}{2}$}}

\newcommand{\del}[2]{\mbox{$\displaystyle\frac{#1}{#2}$}}
\newcommand{\ppd}[2]{\del{\partial {#1}}{\partial{#2}}}
\newcommand{\der}[2]{\del{\de #1}{\de #2}}
\newcommand{\de}[1]{\,{\mathrm d}#1}

\newcommand{\Etot}{\Pi}
\newcommand{\Eint}[1]{\Etot\lay{#1}_{\mathrm{int}}}
\newcommand{\Eext}[1]{\Etot\lay{#1}_{\mathrm{ext}}}
\newcommand{\Eexte}[2]{\Etot\lay{#1}_{\mathrm{ext},#2}}
\newcommand{\Einte}[2]{\Etot\lay{#1}_{\mathrm{int},#2}}

\newcommand{\E}[1]{\mathrm{E}\lay{#1}}%
\newcommand{\G}[1]{\mathrm{G}\lay{#1}}%
\newcommand{\A}[1]{A\lay{#1}}%
\newcommand{\I}[1]{I\lay{#1}}%
\newcommand{\As}[1]{A_\mathrm{s}\lay{#1}}%
 
\newcommand{\Le}{L}
\newcommand{\be}{b} 

\newcommand{\Mu}{\M{u}}
\newcommand{\MS}{\M{E}}
\newcommand{\MN}[1]{\M{N}_{#1}}
\newcommand{\Mde}[2]{\Md^{(#1)}_{#2}}

\newcommand{\el}{e}
\newcommand{\numel}{n^\mathrm{\el}}
\newcommand{\nnode}{n^\mathrm{n}}

\newcommand{\trn}{{\sf ^T}}        % Transpose
\newcommand{\eL}[2]{L_{#1}\lay{#2}}

\newcommand{\ite}[1]{{^{#1}}}
\newcommand{\MKt}{\M{K}_\mathrm{t}}
\newcommand{\Mfint}{\M{f}_\mathrm{int}}
\newcommand{\Mfinte}[1]{\M{f}_{\mathrm{int},#1}}
\newcommand{\Mfext}{\M{f}_{\mathrm{ext}}}
\newcommand{\Mfexte}[2]{\M{f}_{\mathrm{ext},#2}\lay{#1}}
\newcommand{\finte}[1]{f_{\mathrm{int},#1}}

\newcommand{\Kt}[1]{K_{\mathrm{t},#1}}
\newcommand{\Md}{\M{d}}
\newcommand{\Mdev}[3]{\Md_{#3}^{\mathrm{#1}(#2)}}

\newcommand{\etaexp}{\eta_\mathrm{exp}}
\newcommand{\etanum}{\eta_\mathrm{num}}
\newcommand{\etaan}{\eta_\mathrm{an}}

\newcommand{\etal}{et~al.}
\newcommand{\Sref}[1]{Section~\ref{#1}}
\newcommand{\Tref}[1]{Table~\ref{#1}}
\newcommand{\Alref}[1]{Algorithm~\ref{#1}}

\title{Numerical model of elastic laminated glass beams under finite strain}

\makenomenclature

\author[1]{Alena Zemanov\'{a}}
\author[1,2]{Jan Zeman}
\author[1]{Michal \v{S}ejnoha}
\affil[1]{Department of Mechanics, Faculty of Civil Engineering,
  Czech Technical University in Prague, Th\'{a}kurova 7, 166 29 Prague
  6, Czech Republic}
\affil[2]{Centre of Excellence IT4Innovations, V\v{S}B-TU Ostrava, 
17.~listopadu 15/2172 708 33 Ostrava-Poruba, Czech Republic}

\begin{document}

\maketitle

\begin{abstract}
Laminated glass structures are formed by stiff layers of glass
connected with a compliant plastic interlayer. Due to their slenderness and
heterogeneity, they exhibit a complex mechanical response that is difficult to
capture by single-layer models even in the elastic range. The
purpose of this paper is to introduce an efficient and reliable finite element
approach to the simulation of the immediate response of laminated glass beams.
It proceeds from a refined plate theory due to \rev{Mau (1973)}, as we treat each
layer independently and enforce the compatibility by the Lagrange multipliers. 
\rev{At the layer level, we adopt the finite-strain shear
deformable formulation of Reissner~(1972) and the numerical framework by
Ibrahimbegovi\'{c} and Frey~(1993).} The resulting system is solved by the
Newton method with consistent linearization. By comparing the model predictions
against available experimental data, analytical methods and two-dimensional
finite element simulations, we demonstrate that the proposed formulation is
reliable and provides accuracy comparable to the detailed two-dimensional finite
element analyzes. As such, it offers a convenient basis to incorporate
more refined constitutive description of the interlayer.
\end{abstract}


\section{Introduction}\label{sec:introduction}

Due to present trends in architecture and photovoltaics, the use of structural
glass \rev{is expanding} from traditional window panes to large-area surfaces,
roof and floor systems, columns or staircases. This interest leads to an
increased emphasis on the safety and the mechanical performance of structural
members made of glass. Perhaps the most popular material system meeting these
criteria is laminated glass. It is a composite structure produced by bonding
multiple layers of glass together with a transparent plastic interlayer,
typically made of polyvinyl butyral (PVB)
foil~\cite{Haldimann:2008:SUG,Sedlacek:1995:GSE}. The interlayer absorbs the
energy impact, thereby resisting the glass penetration, and keeps the layers of
glass bonded when fractured.

Laminated glass units exhibit a complex mechanical response, as a consequence of
their heterogeneity. Namely, the contrast in elastic properties between glass
and the interlayer typically exceeds three orders of magnitude, which renders
classical laminate theories inapplicable since the glass layers deform mainly
due to bending, whereas the interlayer experiences a pure shear. Moreover, PVB is a
viscoelastic material exhibiting a high sensitivity to temperature changes,
e.g.~\cite{Bennison:1999:FLB}. Finally, glass structures are very slender and
must be analyzed using geometrically non-linear theories.

In practical applications, the behavior of laminated glass units is often
approximated by an equivalent, geometrically linear, single-layer elastic
system. According to the performance of the interlayer, we distinguish the
layered case, in which the structure responds as an assembly of independent
layers, and the monolithic model with thickness equal to the combined thickness
of glass layers and interlayer. This approach was pioneered by experimental
studies of Behr \etal~\cite{Behr:1985:LGU}, who demonstrated that the degree of
coupling due to the interlayer is significant at room temperatures and ceases at
elevated temperatures. This was extended later in~\cite{Behr:1993:SBA} to
quantify the validity of the monolithic approximation in terms of temperature
range and load duration. Norville~\etal~\cite{Norville:1998:BSL} further refined
these results by studying the shear coupling around the transition temperature
of the interlayer and demonstrated that the performance of laminated units
exceeds the layered limit even above the transition point.
Vallabhan~\etal~\cite{Vallabhan:1987:SLG} introduced a notion of the strength
factor as a ratio between the maximal principal strength in the equivalent
monolithic unit and the laminated system and demonstrated that it can reach
values smaller than one for cases of practical interest. They attributed this to
the effects of geometrical non-linearity which become significant earlier for
the laminated units than for the monolithic ones. These developments have been
recently put on a rigorous basis by Galuppi and
Royer-Carfagni~\cite{Galuppi:2012:ETL}, who derived explicit variationally-based
formulas for deflection- and stress-equivalent thicknesses of monolithic
systems.

Apart from experimental results, the validity of simplified models was assessed
by several analytical studies. When restricting our attention to planar glass
beams, the first study is due to Hooper~\cite{Hooper:1973:BAL}, who derived a
three-layer model under small deflections and presented the solution for the
four point bending setup with different duration of loading and ambient
temperatures. Later, A\c{s}\i{}k and Tescan~\cite{Asik:2005:MMB} extended this
model to account for large deflections and demonstrated that they significantly
contribute to the overall response when the normal forces start to develop, see
also \Sref{sec:examples} for a concrete example. In the linear setting,
Ivanov~\cite{Ivanov:2006:AMO} proposed a procedure for thickness optimization of
triplex glasses and Foraboschi~\cite{Foraboschi:2007:BFS} demonstrated that the
empirical rules proposed by Behr~\etal~\cite{Behr:1993:SBA} may lead to unsafe
designs. Recently, Schultze~\etal~\cite{Schultze:2012:ALG} performed an
experimental-analytical study into the response of the laminated structures used
in photovoltaic applications under three-point bending.

Even though the analytical approaches give valuable insight into the behavior of
laminated glass structures, they still suffer from two major limitations. First,
the only closed-form solutions we are aware of hold only in the absence of
membrane effects. In the opposite case, the governing differential equations
have to be discretized anyhow and the discrete problem is often solved using the
problem-specific procedures that experience convergence problems,
e.g.~\cite{Asik:2003:LGP}. Second, the analytical approaches require the
interlayer to be replaced by an equivalent elastic material, with properties
adjusted to the loading duration and ambient temperature. As shown by Galuppi
and Royer-Carfagni~\cite{Galuppi:2012:LBV}, this approach leads to significant
errors in local stresses and strains.

With these limitations in mind, we proposed in~\cite{Zemanova:2008:SNM} a
numerical approach to the analysis of laminated glass beams based on a refined
laminate theory due to Mau~\cite{Mau:1973:RLP}. In this framework, the structure
is seen as an assembly of shear-deformable layers with independent kinematics,
and the inter-layer interaction is enforced by the Lagrange multipliers.
\rev{Such an approach may appear costly, as it increases the number of
unknowns and tends to produce badly conditioned systems of equations, but it
offers several advantages due to the modular format. Namely, each layer can be
discretized by an appropriate type of finite elements, and different
constitutive models can be used at the layer level (even in the multi-scale
setting~\cite{Sejnoha:1996:MMU}). From the computational perspective, the
resulting system is ideally suited for the deployment of efficient duality-based iterative
solvers~\cite{Kruis:2002:SLP} that can also be extended to account for
inter-layer delamination, e.g.~\cite{Kruis:2008:RMI,Roubicek:2013:DAC}.}

In this paper, we extend our previous
results~\cite{Zemanova:2008:SNM}, valid in small strains, to the finite-strain
regime. For this purpose, each layer is modeled by the Reissner beam
theory~\cite{Reissner:1972:ODFSBT}, briefly summarized in \Sref{sec:formulation}
in a variational format. Discretization in \Sref{sec:discretization} is
based on a robust finite element formulation proposed by Ibrahimbegovi\'{c} and
Frey in~\cite{Ibrahimbegovic:1993:FEA} and later applied, e.g., to discrete
materials modeling~\cite{Ibrahimbegovic:2003:MMD}, or optimal control and design
of structures~\cite{Ibrahimbegovic:2004}. In \Sref{sec:governing_equations}, we
derive the discretized system arising from the optimality conditions of the
associated constrained optimization problem, and perform its solution by the
Newton method. Accuracy of the implementation is examined in \Sref{sec:examples}
by verifying our results against data presented by A\c{s}\i{}k and
Tescan~\cite{Asik:2005:MMB}. To make the paper self-contained, in
\Aref{app:sensitivity_analysis} we collect the details on tangent operators
needed when implementing the Newton method. Note that this paper is focused on
the immediate elastic response of laminated structures. The effect of
\rev{temperature}-dependent viscous properties of the interlayer will be
discussed independently and in more details in a forthcoming publication.

The following nomenclature is used in the text. Scalar quantities are denoted by
lightface letters, e.g. $a$, and the bold letters are reserved for matrices,
e.g. $\M{a}$ or $\M{A}$. $\M{A}\trn$ standardly stands for the matrix transpose
and $\M{A}^{-1}$ for the matrix inverse. The subscript in parentheses, e.g.
$a\lay{i}$, is used to emphasize that the variable $a$ is associated with the
$i$-th layer.

\section{Model formulation}\label{sec:formulation}

\begin{figure}[ht]
\centerline{
 \includegraphics{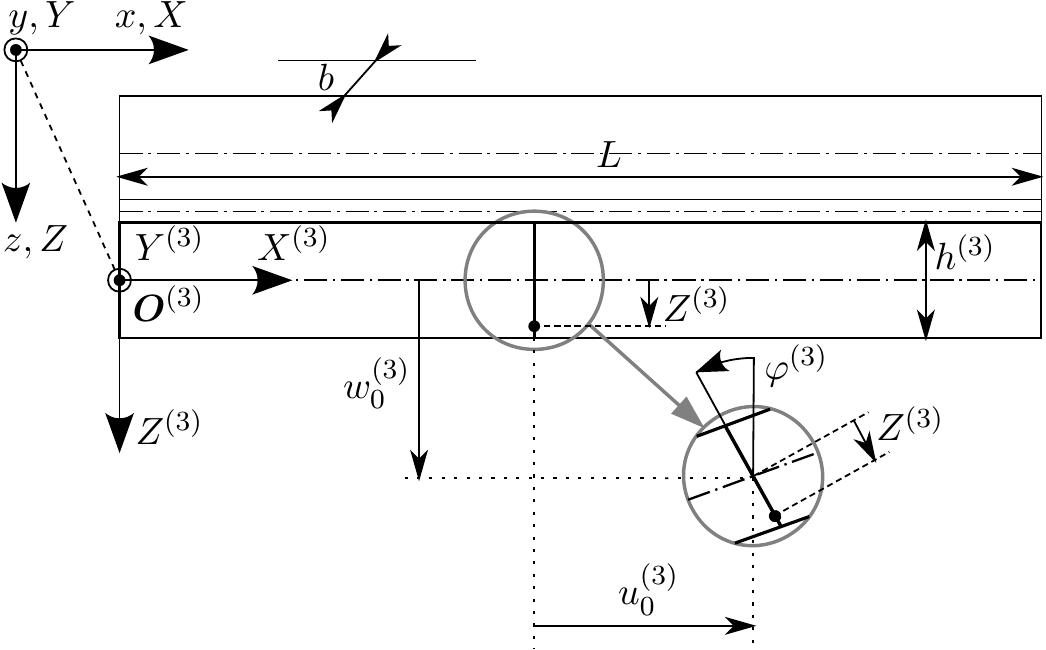} 
}
\caption{Kinematics of a cross section of the bottom layer of a laminated
beam~($i=3$).}
\label{fig:lam_beam}
\end{figure}

In our setting, a glass beam is considered as an assembly of
three beams of identical length $\Le$, with the cross-section dimensions $\be \times \h{i}$.
\nomenclature{$\Le$}{Lend of the beam}%
\nomenclature{$\be$}{Beam width}%
\nomenclature{$\h{i}$}{Height of the $i$-th beam}%
\nomenclature{$i$}{Layer index}%
Each layer is considered to behave according to the Reissner finite-strain beam
theory~\cite{Reissner:1972:ODFSBT}, i.e. we assume that cross-sections remain
planar, but not necessarily perpendicular to the deformed beam curve, and that
the distance of a point at the cross-section from the centerline remains
unchanged. For the $i$-th layer, the coordinates of a point in the deformed
configuration can be determined as, \Fref{fig:lam_beam},
\begin{subequations}\label{eq:coo_def}
\begin{align}
\x{i}(\X{i}, \Z{i}) 
& = 
O\lay{i}_X 
+ 
\X{i} + \uc{i}(\X{i}) 
+
\sin\bigl( 
  \rot{i}(\X{i}) 
\bigr)
\Z{i}, 
\\ 
\z{i}(\X{i},\Z{i}) 
& 
= 
O\lay{i}_Z 
+ 
\wc{i}(\X{i}) 
+
\cos\bigl(
  \rot{i}(\X{i})
\bigr) 
\Z{i},
\end{align}
\end{subequations}
\nomenclature{$a\lay{i}$}{Quantity $a$ related to the $i$-th layer}%
\nomenclature{$\rot{i}$}{Rotation}%
\nomenclature{$\x{i}$}{Position in the deformed configuration}%
\nomenclature{$\X{i}$}{Position in the original configuration}%
\nomenclature{$\z{i}$}{Position in the deformed configuration}%
\nomenclature{$\X{i}$}{Position in the original configuration}%
\nomenclature{$\uc{i}$}{Centerline displacements}%
\nomenclature{$\wc{i}$}{Centerline displacements}%
where $O\lay{i}_X$ and $O\lay{i}_Z$ stand for the coordinates of the beam
origin, $\X{i}$ is the ordinate of a cross-section, $\uc{i}$ and~$\wc{i}$ are
centerline displacements measured in the global coordinate system, $\rot{i}$ is
the rotation of the cross-section, and $\Z{i}$ is the coordinate measured along
the cross-section. The inter-layer interaction is ensured via the geometric
continuity conditions at the interfaces between the layers~($i=1,2$)
\begin{subequations}
\begin{align}\label{eq:continuity2}
\x{i}( \X{i}, \half\h{i} ) 
-
\x{i+1}(\X{i},-\half\h{i+1}) 
& = 0,
\\
\z{i}(\X{i},\half\h{i}) 
- 
\z{i+1}(\X{i},-\half\h{i+1}) 
& = 0.
\end{align}
\end{subequations}

As demonstrated, e.g. by Ibrahimbegovi\'{c}
and~Frey~\cite{Ibrahimbegovic:1993:FEA} and Irschik
and~Gerstmayr~\cite{Irschik:2009:CMBDR}, the Reissner beam theory can be
consistently derived from the continuum framework by the Biot-type strain
tensors for the kinematics parametrization~\eqref{eq:coo_def}. To this purpose,
we use the deformation gradient in the form
\begin{equation}
\MF{i}(\X{i}, \Z{i})
=
\begin{bmatrix}
  \ppd{\x{i}}{\X{i}} &
  \ppd{\x{i}}{\Z{i}} \\
  \ppd{\z{i}}{\X{i}} &
  \ppd{\z{i}}{\Z{i}}
\end{bmatrix}(\X{i},\Z{i})
\label{eq:def_grad},
\end{equation}
\nomenclature{$\MF{i}$}{Deformation gradient in the $i$-th layer}%
with individual entries provided by
\begin{subequations}%
\begin{align}
\F{i}{11} & =
1 
+ 
\der{\uc{i}(\X{i})}{\X{i}}
+
\cos\bigl( \rot{i}(\X{i}) \bigr) 
\der{\rot{i}(\X{i})}{\X{i}} \Z{i},
\\
\F{i}{12} & = 
\sin \bigl(\rot{i}(\X{i})\bigr), 
\\ 
\F{i}{21} & = 
\der{\wc{i}(\X{i})}{\X{i}}
-
\sin\bigl( \rot{i}(\X{i}) \bigr)
\der{\rot{i}(\X{i})}{\X{i}} \Z{i},
\\
\F{i}{22} & = 
\cos \bigl( \rot{i}(\X{i}) \bigr).
\end{align}
\end{subequations}
The deformation gradient $\MF{i}$ admits a multiplicative decomposition
\begin{eqnarray}
\MF{i}(\X{i}, \Z{i}) = \MR{i}(\X{i}) \MU{i}(\X{i}, \Z{i}),
\end{eqnarray}
\nomenclature{$\MR{\bullet}$}{Rotation tensor}%
\nomenclature{$\MU{\bullet}$}{Stretch tensor}%
where the rotation matrix of the $i$-th layer is given by, recall
\Fref{fig:lam_beam}, 
\begin{equation}
\MR{i}
=
\begin{bmatrix}
\cos \bigl( \rot{i}(\X{i}) \bigr) & \sin \bigl( \rot{i}(\X{i}) \bigr) \\ 
-\sin \bigl( \rot{i}(\X{i}) \bigr) & \cos \bigl( \rot{i}(\X{i}) \bigr)
\end{bmatrix},
\end{equation}
and $\MU{i}$ \rev{follows} from the inverse relation
\begin{eqnarray}
\MU{i}(\X{i}, \Z{i}) 
= 
\bigl[ 
  \MR{i}(\X{i}) 
\bigr]^{-1} 
\MF{i}(\X{i},\Z{i}),
\end{eqnarray}
that provides 
\begin{subequations}
\begin{align}
\U{i}{11} & 
=  
\cos\bigl( \rot{i}(\X{i}) \bigr)
\Bigl(
1 + \der{\uc{i}(\X{i})}{\X{i}}
\Bigr)
- 
\sin\bigl( \rot{i}(\X{i}) \bigr)
\der{\wc{i}(\X{i})}{\X{i}}
+
\der{\rot{i}(\X{i})}{\X{i}}\Z{i},
\\
\U{i}{12} 
& = 0,
\\
\U{i}{21} 
& = 
\sin \bigl( \rot{i}(\X{i}) \bigr) 
\Bigl( 
  1 + \der{\uc{i}(\X{i})}{\X{i}}
\Bigr) 
+ 
\cos \bigl(\rot{i}(\X{i}) \bigr) 
\der{\wc{i}(\X{i})}{\X{i}}
, \\
\U{i}{22} & = 1.
\end{align}
\end{subequations}
Employing the definition of the Biot-type strain tensor,
e.g.~\cite[Eq.~(24.27)]{Jirasek:2002:IAS},
\begin{equation}
\MH{i}(\X{i}, \Z{i})
=
\MU{i}(\X{i}, \Z{i})
-
\M{I},
\end{equation}
\nomenclature{$\MH{\bullet}$}{Biot strain tensor}%
the non-zero strain components are given by
\begin{align}
\Ht{i}{11}(\X{i}, \Z{i})
= 
\Sc{i}(\X{i}) + \Kc{i}(\X{i})\Z{i},
&&
\Ht{i}{21}(\X{i}) = \Gc{i}(\X{i}),
\label{eq:H}
\end{align}
\nomenclature{$\Sc{\bullet}$}{Reissner Normal strain}%
\nomenclature{$\Kc{\bullet}$}{Reissner Pseudo-curvature}%
\nomenclature{$\Gc{\bullet}$}{Reissner Shear stress}%
where $\Sc{i}$, $\Kc{i}$ and $\Gc{i}$ denote the generalized normal strain,
pseudo-curvature and shear strain introduced by
Reissner~\cite{Reissner:1972:ODFSBT}:
\begin{subequations}\label{eq:generalized_measures}
\begin{align}
\Sc{i}
& = 
E( \uc{i}, \wc{i}, \rot{i} )
\nonumber \\
& =
\cos \bigl( \rot{i}(\X{i}) \bigr)
\Bigl( 
  1 + \der{\uc{i}(\X{i})}{\X{i}}
\Bigr)
- 
\sin \bigl( \rot{i}(\X{i}) \bigr)
\der{\wc{i}(\X{i})}{\X{i}}
-
1,
\\
\Gc{i} 
& = 
\Gamma( \uc{i}, \wc{i}, \rot{i} )
\nonumber \\
& =
\sin \bigl( \rot{i}(\X{i}) \bigr) 
\Bigl( 
  1 + \der{\uc{i}(\X{i})}{\X{i}}
\Bigr)
+ 
\cos \bigl( \rot{i}(\X{i}) 
\der{\wc{i}(\X{i})}{\X{i}},
\\
\Kc{i} 
& =  
K( \uc{i}, \wc{i}, \rot{i} )
=
\der{\rot{i}(\X{i})}{\X{i}}.
\end{align}
\end{subequations}
It is useful to express these relations in a compact form 
\begin{align}
\Mu\lay{i}( \X{i} )
= 
\begin{bmatrix}
\uc{i} \\ \wc{i} \\ \rot{i}
\end{bmatrix}
( \X{i} )
, &&
\MS\lay{i}( \X{i} )
=
\begin{bmatrix}
\Sc{i} \\ \Gc{i} \\ \Kc{i}
\end{bmatrix}
( \X{i} ),
\end{align}
\nomenclature{$\Mu$}{Matrix of generalized centerline displacements}%
\nomenclature{$\MS$}{Matrix of the Reissner strain measures}%
and denote by $\MS( \Mu )$ a mapping assigning the generalized strain
measures to the generalized centerline displacements $\Mu$ according to
\Eref{eq:generalized_measures}.

The model is completed by specifying the energy functionals associated
with the deformation of the laminated beam. In particular, the internal energy
of the $i$-th layer is provided by~\cite{Ibrahimbegovic:1993:FEA}
\begin{align}\label{eq:energy_internal}
\Eint{i} \bigl( \MS\lay{i} \bigr) 
 = &  \\
&
\half
\int_0^L
 	\E{i}\A{i} \bigl( \Sc{i}(\X{i}) \bigr)^2
 	+
 	\G{i}\As{i} \bigl( \Gc{i}(\X{i}) \bigr)^2
 	+
 	\E{i}\I{i} \bigl( \Kc{i}(\X{i}) \bigr)^2
\de \X{i},\nonumber
\end{align}
where $\E{i}$ and $\G{i}$ denote the Young and shear moduli, and 
$\A{i}=\be \h{i}$, $\As{i} = \frac{5}{6} \A{i} $ and $\I{i}=\frac{1}{12} \be
(\h{i})^3$ stand for the the cross-section area, effective shear area, and the
second moment of area, respectively.
\nomenclature{$\Eint{\bullet}$}{Internal energy of layer $\bullet$}%
\nomenclature{$\E{\bullet}$}{Young's modulus of $\bullet$ layer}%
\nomenclature{$\G{\bullet}$}{Shear modulus of $\bullet$ layer}%
\nomenclature{$\A{\bullet}$}{Area of $\bullet$ layer}%
\nomenclature{$\I{\bullet}$}{Moment of inertia of $\bullet$ layer}%
\nomenclature{$L$}{Length of the beam}%
The external energy due to loading acting at the $i$-th layer assumes the form
\begin{align}\label{eq:energy_external}
\Eext{i}\bigl( \Mu\lay{i} \bigr) 
= 
- 
\int_0^L \wc{i}(\X{i}) \overline{f}_Z\lay{i}(\X{i})\de \X{i}
-
\sum_{p=1}^{n^{\mathrm{p}(i)}}
\wc{i}(\X{i}_p)
\overline{F}\lay{i}_Z(\X{i}_p),
\end{align}
\nomenclature{$\Eext{\bullet}$}{External energy of $\bullet$}%
\nomenclature{$p$}{Index used to number loads}%
where, for simplicity, we assume that the structure is subjected to the
distributed loading with intensity $\overline{f}\lay{i}_Z$ and to
$n^{\mathrm{p}(i)}$ concentrated forces $\overline{F}\lay{i}_Z$ acting
at $\X{i}_p$. The total energy of the $i$-th layer is then given by 
\begin{equation}\label{eq:layer_energy_def}
\Etot\lay{i}\bigl( \Mu\lay{i} \bigr)
=
\Eint{i}
\bigl( \MS( \Mu\lay{i} ) \bigr)
+
\Eext{i}
\bigl( \Mu\lay{i} \bigr), 
\end{equation}
\nomenclature{$\Etot$}{Total energy}%
and at the level of the whole structure it reads
\begin{equation}
\Etot
\bigl(
	\Mu\lay{1}, \Mu\lay{2}, \Mu\lay{3}
\bigr)
=
\sum_{i=1}^3
\Etot\lay{i}
\bigl( \Mu\lay{i} \bigr).
\end{equation}

\section{Finite element discretization}\label{sec:discretization}
\begin{figure}[ht]
\centerline{
 \includegraphics{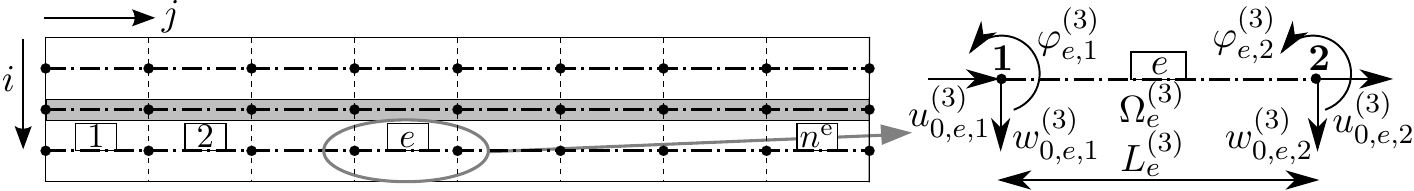} 
}
\caption{Scheme of discretization and element degrees of freedom.}
\label{fig:lam_beam_discr}
\end{figure}

It is convenient for the numerical treatment to discretize all layers
identically with $\numel$ two-node elements per each layer; by
$\Omega\lay{i}_\el$ we denote the $\el$-th element of the $i$-th layer. The displacement fields at the
element level are approximated as
\nomenclature{$\el$}{Element index}%
\nomenclature{$\numel$}{Number of elements per layer}%
\nomenclature{$\Omega\lay{i}_\el$}{$\el$-th element of the $i$-th layer}%
\begin{equation}\label{eq:approximate_fields}
\uc{i}(x)
\approx 
\MN{\el}(x)
\Mdev{u}{i}{\el},
\;
\wc{i}(x)
\approx 
\MN{\el}(x)
\Mdev{w}{i}{\el},
\;
\rot{i}(x)
\approx 
\MN{\el}(x)
\Mdev{\varphi}{i}{\el}
\text{ for } 
x \in \Omega\lay{i}_\el,
\end{equation}
where $\MN{\el}$ is the matrix of piecewise linear basis functions, and e.g.
$\Mdev{\varphi}{i}{\el} = \bigl[\rot{i}_{\el,1}, \rot{i}_{\el,2} \bigr]\trn$
collects the nodal rotations $\rot{i}$, cf.~\Fref{fig:lam_beam_discr}.
\nomenclature{$\MN{\el}$}{Matrix of basis functions for $\el$-th element}%
\nomenclature{$\Md{\bullet}{\el}{i}$}{Nodal unknowns for $\bullet$ at element
$\el$ and layer $i$}%

Inserting the approximate fields~\eqref{eq:approximate_fields} into the energy
functionals~\eqref{eq:layer_energy_def} results in
\begin{eqnarray}
\Eint{i}( \Mu\lay{i} )
\approx 
\sum_{\el=1}^{\numel}
\Einte{i}{\el}( \Mde{i}{\el} ),
&&
\Eext{i}( \Mu\lay{i} )
\approx
\sum_{\el=1}^{\numel}
\Eexte{i}{\el}(\Mde{i}{\el})
=
-
\sum_{\el=1}^{\numel}
\Mde{i}{\el}
\trn 
\Mfexte{i}{e},
\end{eqnarray}
\nomenclature{$\Einte{i}{\el}$}{Element internal energy}% 
\nomenclature{$\Eext{i}{\el}$}{Element external energy}% 
\nomenclature{$\Mde{i}{\el}$}{Total nodal unknowns for respective element}%
where the matrix $\Mde{i}{\el}$ collects the nodal unknowns, 
\begin{eqnarray}
\Mde{i}{\el}
=
\begin{bmatrix}
\uce{i}{\el,1} & \wce{i}{\el,1} & \rote{i}{\el,1} & 
\uce{i}{\el,2} & \wce{i}{\el,2} & \rote{i}{\el,2} 
\end{bmatrix}\trn,
\end{eqnarray}
recall \Fref{fig:lam_beam_discr}, and $\Mfexte{i}{e}$ stores the corresponding
generalized nodal forces. As demonstrated in detail by Ibrahimbegovi\'{c} and
Frey~\cite{Ibrahimbegovic:1993:FEA}, the normal and shear locking even for the
non-linear kinematics can be suppressed by adopting the selective one-point
integration of the corresponding terms in~\eqref{eq:energy_internal}. Thus, the
approximate generalized strain measures~\eqref{eq:generalized_measures} are
taken as element-wise constants:
\begin{subequations}\label{eq:element_measures}
\begin{align}
\Sc{i}_\el(\Mde{i}{\el})
& = 
\frac{1}{\eL{\el}{i}}
\left( 
\eL{\el}{i} 
+ 
\Duc{i}{\el} \right)
\cos \beta\lay{i}_\el 
- 
\frac{1}{\eL{\el}{i}}
\Dwc{i}{\el}
\sin \beta\lay{i}_\el
- 1,
\\
\Gc{i}_\el(\Mde{i}{\el}) 
& = 
\frac{1}{\eL{\el}{i}}
\left( \eL{\el}{i} + \Duc{i}{\el} \right)
\sin \beta\lay{i}_\el
+ \frac{1}{\eL{\el}{i}}
\Dwc{i}{\el}
\cos \beta\lay{i}_\el,
\label{eq:SFe}
\\
\Kc{i}_\el(\Mde{i}{\el}) 
& = 
\frac{\Drot{i}{\el}}{\eL{\el}{i}},
\end{align}
\end{subequations}
\nomenclature{$\beta\lay{i}_\el$}{Mean element rotation}%
with, e.g., $\Drot{i}{\el} = \rot{i}_{\el,2} - \rot{i}_{\el,1}$, 
$\beta\lay{i}_\el = \half(\rot{i}_{\el,1} + \rot{i}_{\el,2})$, and
$\eL{\el}{i}$ denoting the element length. The contribution of the $\el$-th
element to the internal energy simplifies to
\begin{equation}\label{eq:Einte}
\Einte{i}{\el}
(\Mde{i}{\el})
=
\half
\left(
\E{i}\A{i} 
\Sc{i}_\el \bigl(\Mde{i}{\el} \bigr)^2
+
\G{i}\As{i} 
\Gc{i}_\el \bigl(\Mde{i}{\el} \bigr)^2
+
\E{i}\I{i} 
\Kc{i}_\el \bigl(\Mde{i}{\el} \bigr)^2
\right)
\eL{\el}{i}.
\end{equation}

The discretization is completed by enforcing the inter-layer compatibility
conditions~\eqref{eq:continuity2} at the element nodes, 
\nomenclature{$\nnode$}{Number of nodes}%
\nomenclature{$j$}{Index used to number nodes}%
\nomenclature{$c\lay{i,i+1}_{\bullet,j}$}{Compatibility condition for $j$-th
node and direction $\bullet$}%
\begin{eqnarray}\label{eq:CC2}
c\lay{i,i+1}_{X,j} = 0, && c\lay{i,i+1}_{Z,j} = 0,
\end{eqnarray}
indexed at the level of a
single layer by $j=1, 2, \ldots, \numel+1$, cf. \Fref{fig:lam_beam}, and with
\begin{subequations}
\begin{align}
c\lay{i,i+1}_{X,j} 
& = 
\uce{i}{j}
-
\uce{i+1}{j}
+
\half ( \h{i}
\sin\rote{i}{j}
+
\h{i+1}
\sin\rote{i+1}{j}
),
\\ 
c\lay{i,i+1}_{Z,j} 
& = 
-\half(\h{i} + \h{i+1})
+
\wce{i}{j} - \wce{i+1}{j}
+
\half(\h{i}\cos\rote{i}{j}
+
\h{i+1} \cos\rote{i+1}{j} ).
\end{align}
\end{subequations}
For later reference, these are introduced in a compact form as
\begin{equation}
\M{c}( \Md ) = \M{0},
\end{equation}
where $\Md = [ \Md\lay{1}, \Md\lay{2}, \Md\lay{3} ]$ is a $9
(\numel+1) \times 1$ column matrix of nodal degrees of freedom and $\M{c}$
collects the $4 (\numel + 1)$ compatibility conditions~\eqref{eq:CC2}.

\section{Governing equations}\label{sec:governing_equations}

The true nodal displacements $\Md^*$ follow from the minimization of the
discretized energy functional
\begin{equation}\label{eq:Eint}
\Etot( \Md )
=
\sum_{i=1}^{3}
\sum_{\el=1}^{\numel}
\Einte{i}{\el}( \Mde{i}{\el} )
+
\Eexte{i}{\el}( \Mde{i}{\el} ),
\end{equation}
subject to both kinematic constraints and compatibility
conditions~\eqref{eq:CC2}. While the kinematic constrains are dealt with by
matrix reduction techniques, e.g.~\cite[Appendix~A]{Jirasek:2002:IAS}, the
compatibility is enforced via the Lagrange
multipliers~\cite{Mau:1973:RLP,Sejnoha:1996:MMU}. This procedure yields the
Lagrangian function in the form
\begin{equation}\label{eq:nonl_lagrangian}
\mathcal{L}( \Md, \M{\lambda} )
=
\Etot( \Md )
+
\M{\lambda}\trn\M{c}( \Md )
=
\Etot( \Md )
+
\sum_{m=1}^{4(\numel+1)}
\lambda_m c_m (\Md)
\end{equation}
\nomenclature{$m$}{Index for Lagrange multipliers}%
where $\M{\lambda}$ is a $4(\numel+1) \times 1$ matrix storing the corresponding
Lagrange multipliers.

The corresponding Karush-Kuhn-Tucker optimality conditions read,
e.g.~\cite[Chapter~14]{Bonnans:2003:NOTPA},
\begin{subequations}\label{eq:opt_conditions}
\begin{align}
\nabla_{\Md} 
\mathcal{L}\left( 
 \Md^*, \M{\lambda}^* 
\right)
=  
\nabla \Etot( \Md^* )
+ 
\nabla \M{c}( \Md^* )\trn 	
\M{\lambda}^*
& =  
\M{0},
\label{eq:opt_conditions_1}
\\
\nabla_{\M{\lambda}} 
\mathcal{L}\left( 
 \Md^*, \M{\lambda}^* 
\right)
=
\M{c}( \Md^* )
& =  
\M{0},
\label{eq:opt_conditions_2}
\end{align} 
\end{subequations}
with $\nabla$ denoting the gradient operator and $\nabla_{\bullet}$ designating
the partial derivative with respect to variable~$\bullet$. These relations
represent a system of non-linear equations, to be solved using the Newton
iterative scheme.

To that end, we assume that displacements at iteration $( \ite{k}\Md,
\ite{k}\M{\lambda} )$ are known and search for the iterative correction in the
form
\begin{equation}\label{eq:r_deltar}
\ite{k+1}\Md
=
\ite{k}\Md
+
\ite{k+1}\delta\Md.
\end{equation}
The values of $\ite{k+1}\delta\Md$ are obtained by means of the linearized
expressions
\begin{subequations}
\begin{align}
\nabla\Etot( \ite{k+1}\Md ) 
& \approx 
\nabla\Etot( \ite{k}\Md )
+
\nabla^2 \Etot( \ite{k}\Md)
\ite{k+1}\delta\Md,
\\
\M{c}( \ite{k+1}\Md )
& \approx 
\M{c}( \ite{k}\Md )
+
\nabla \M{c}( \ite{k}\Md )
\ite{k+1}\delta\Md,
\\
\nabla \M{c}( \ite{k+1}\Md )
& \approx 
\nabla \M{c}( \ite{k}\Md )
+
\nabla^2 \M{c}( \ite{k}\Md )
\ite{k+1}\delta\Md,
\end{align}
\end{subequations}
which, when introduced into the optimality
conditions~\eqref{eq:opt_conditions_1} and~\eqref{eq:opt_conditions_2}, result
in a linear system of equations, cf.~\cite{Bonnans:2003:NOTPA}
and~\cite{Zemanova:2008:SNM},
\begin{equation}\label{eq:system}
\begin{bmatrix}
 \ite{k}\M{K} & \ite{k}\M{C}\trn \\
 \ite{k}\M{C} & \M{0}
\end{bmatrix}
\begin{bmatrix}
 \ite{k+1} \delta \Md \\
 \ite{k+1} \M{\lambda}
\end{bmatrix}
=
-
\begin{bmatrix}
 \ite{k}\Mfint - \Mfext \\
 \ite{k}\M{c} 
\end{bmatrix}
.
\end{equation}
Here, we employ the short-hand notation
\begin{subequations}
\begin{align}
\ite{k}\M{K} 
& =  
\nabla^2 \Etot( \ite{k}\Md )
+
\sum_{m=1}^{4(\numel+1)}
\ite{k}\lambda_m
\nabla^2 c_m( \ite{k}\Md )
=
\MKt( \ite{k}\Md ) 
+ 
\M{K}_\lambda( \ite{k}\Md, \ite{k}\M{\lambda}),
\\
\ite{k}\M{C} 
& =  
\nabla\M{c}( \ite{k}\Md ),
\\
\ite{k}\Mfint - \Mfext
& =  
\nabla \Etot( \ite{k}\Md ),
\\
\rev{
\ite{k}\M{c}} & = \rev{\M{c}(\ite{k}\Md).}
\end{align}
\end{subequations}
It follows from the specific form of the energy function~\eqref{eq:Eint} that
the stiffness matrix~$\MKt$ and the matrices of internal $\Mfint$ and external
forces $\Mfext$ exhibit the block structure
\begin{equation}\label{eq:matrices_block_diagonal}
\MKt( \Md )
=
\begin{bmatrix}
\MKt\lay{1}( \Md\lay{1} ) \\ 
& 
\MKt\lay{2}( \Md\lay{2} ) \\
&&  
\MKt\lay{3}( \Md\lay{3} )
\end{bmatrix}
, \quad
\Mfint
=
\begin{bmatrix}
 \Mfint\lay{1} \\
 \Mfint\lay{2} \\
 \Mfint\lay{3}
\end{bmatrix}
, \quad
\Mfext
=
\begin{bmatrix}
 \Mfext\lay{1} \\
 \Mfext\lay{2} \\
 \Mfext\lay{3} 
\end{bmatrix}.
\end{equation}
For the $i$-th layer, these are obtained by the assembly of contributions of
the $e$-th element~\cite{Jirasek:2002:IAS}, provided by
\begin{align}\label{eq:internal_forces}
\M{f}_{\mathrm{int},e}\lay{i}
=
\ppd{\Etot_{\mathrm{int},e}\lay{i}}{\Mde{i}{\el}}, 
&&
\M{K}_{\mathrm t,e}\lay{i} 
=
\ppd{^2\Etot_{\mathrm{int},e}\lay{i}}{\Mde{i}{\el}{}^2} 
=
\ppd{\Mfinte{\el}\lay{i}}{\Mde{i}{\el}}.
\end{align} 
In order to keep the paper self-contained, explicit expressions for the matrices
needed to set up the linearized system~\eqref{eq:system} are summarized
in~\Aref{app:sensitivity_analysis}. Termination of the iterative
process~\eqref{eq:system} is driven by two
residuals~\cite[Section~14.1]{Bonnans:2003:NOTPA}
\begin{align}\label{eq:residuals}
\ite{k}\eta_1 
= 
\frac{%
\| \ite{k}\Mfint - \Mfext + \ite{k}\M{C}\trn
\ite{k}\M{\lambda} \|_2
}{%
\| \Mfext \|_2
},
&&
\ite{k}\eta_2 
= 
\frac{%
\| \ite{k} \M{c} \|_2
}{%
\min_{i} \h{i}
}, 
\end{align}
quantifying the validity of the first-order optimality
conditions~\eqref{eq:opt_conditions} \rev{that are related to equilibrium
conditions for all three layers and the displacement compatibility at
inter-layer interfaces, respectively.} This provides the last component of the
non-linear iterative solver, the implementation of which is outlined in
\Alref{fig:impl_NR}.

\begin{algorithm}[h]
 \KwData{initial displacement $\ite{0}\Md$, tolerances $\epsilon_1$ and
 $\epsilon_2$}
 \KwResult{$\Md^*$, $\M{\lambda}^*$}
 $k \leftarrow 0, \ite{0}\M{\lambda} \leftarrow \M{0}$, 
 assemble $\ite{k}\Mfint$, $\ite{k}\M{c}$ and $\ite{k}\M{C}$\\ 
 \While{$(\ite{k}\eta_1 > \epsilon_1)$ or $(\ite{k}\eta_2 > \epsilon_2)$}{%
 assemble $\ite{k}\M{K}$ \\
 solve for $(\ite{k+1}\delta \Md, \ite{k+1}\M{\lambda})$ from \Eref{eq:system}\\ 
 $\ite{k+1} \Md \leftarrow \ite{k}\Md + \ite{k+1}\delta \Md$\\
 assemble $\ite{k}\Mfint$, $\ite{k}\M{c}$ and $\ite{k}\M{C}$\\ 
 $k\leftarrow k+1$ \\
 } 
 $\Md^* \leftarrow \ite{k}\Md$, $\M{\lambda}^* \leftarrow \ite{k}
 \M{\lambda}$
 \caption{Conceptual implementation of the Newton method.}
 \label{fig:impl_NR}
\end{algorithm}

\section{Examples}\label{sec:examples}

In this section, the proposed finite element formulation is verified and
partially validated against two benchmarks after A\c{s}\i{}k and
Tescan~\cite{Asik:2005:MMB}, involving simply supported,
\Sref{sec:simply_supported}, and fixed-end, \Sref{sec:fixed_end}, three-layer
beams subjected to three-point bending. The comparison is based on the
centerline deflections and \rev{the extreme normal (at top and bottom surfaces)
and shear stresses at the $i$-th layer. These quantities} are obtained from the
element-wise constant strain measures in the form
\rev{%
\begin{align}\label{eq:stresses}
S\lay{i}_{\el,\mathrm{top}}
=
\E{i}
\left( 
\Sc{i}_\el 
-
\half \Kc{i}_\el \h{i}
\right),
&&
S\lay{i}_{\el,\mathrm{bot}}
=
\E{i}
\left( 
\Sc{i}_\el 
+
\half \Kc{i}_\el \h{i}
\right),
&&
T\lay{i}_\el
= 
\G{i} \Gc{i}_\el,
\end{align}}
cf.~\Eref{eq:H} and \Eref{eq:element_measures}, and extrapolated to nodes by the
least-square method assuming the piecewise linear distribution of stresses,
e.g.~\cite{Hinton:1974:LGS}. Accuracy of our results against reference
experimental data is quantified by
\begin{equation}
\rev{%
\etaexp
=
\frac{
(\bullet) - (\bullet)_\mathrm{exp} 
}{
(\bullet)_\mathrm{exp}
}}
,
\end{equation}
and a similar approach is adopted for reference data obtained by analytical
model~(an) or two-dimensional finite element simulations~(num), both
from~\cite{Asik:2005:MMB}.

Results of the finite-strain formulation are complemented with the small-strain
version~\cite{Zemanova:2008:SNM} that corresponds to the first iteration of
\Alref{fig:impl_NR}\rev{, and the linearized strain measures given by
\begin{align*}
\Sc{i}_\el(\Mde{i}{\el})
= 
\frac{\Duc{i}{\el}}{\eL{\el}{i}}
, &&
\Gc{i}_\el(\Mde{i}{\el}) 
= 
\beta\lay{i}_\el
+
\frac{\Dwc{i}{\el}}{\eL{\el}{i}}
, &&
\Kc{i}_\el(\Mde{i}{\el}) 
= 
\frac{\Drot{i}{\el}}{\eL{\el}{i}},
\end{align*}
recall \Eref{eq:element_measures}. The corresponding stresses follow from the
same relation as for the non-linear model, \eqref{eq:stresses}.} For
completeness, we also provide the response of the equivalent monolithic beam of
total thickness $(\h{1} + \h{2} + \h{3})$ and the layered beam corresponding to
two independent layers of thicknesses $\h{1}$ and $\h{3}$, \rev{under the
assumption of geometric linearity}.\footnote{All results in the present section
are reproducible with MATLAB$^\text{\textregistered}$ scripts available at
\url{http://arxiv.org/abs/1303.6314}.} Note that to avoid confusion with the
terminology used for composite structures~\cite{Mau:1973:RLP}, the layered
approximation will \rev{be} referred to as \emph{two-layer} and the present
formulation is denoted as \emph{refined} in what follows.

\subsection{Simply supported beam}\label{sec:simply_supported}

\begin{figure}[h]
\begin{tabular}{ll}
\includegraphics[width=.49\textwidth]{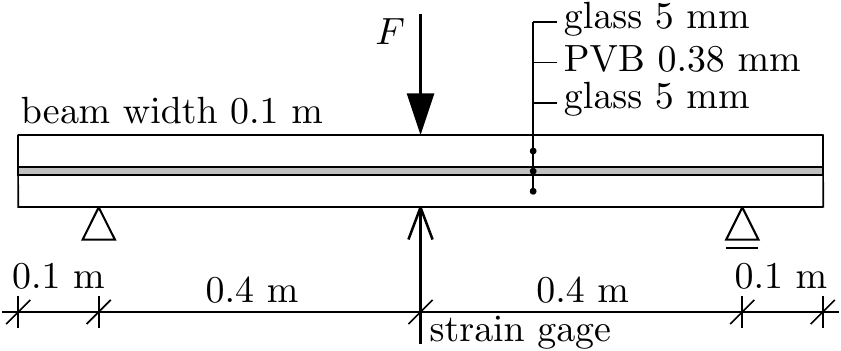} & 
\includegraphics[width=.49\textwidth]{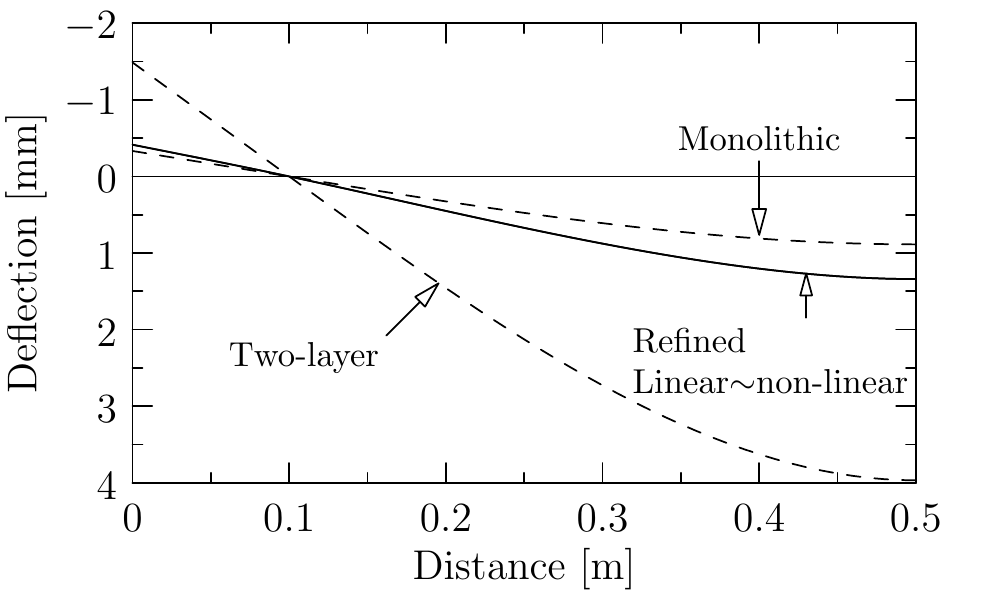} 
\\[-5mm]
(a) & (b)
\end{tabular}
\caption{Simply supported three-layer beam; (a)~experiment setup and
(b)~centerline deflections for $F = 50$~N.}
\label{fig:simply_supported}
\end{figure}

The first example concerns a simply supported beam with a span of $0.8$~m and
the total length of $1$~m, with a symmetric layer setup of 5--0.38--5~mm in
thickness. The structure is subject to a concentrated force at the mid-span
ranging from 50 to 200~N and the quantities of interest are registered at the
bottom layer, cf.~\Fref{fig:simply_supported}(a). The material data for
individual layers appear in
\Tref{tab:simply_supported_data}.\footnote{\rev{Since PVB layer exhibits
viscoelastic and temperature-dependent behavior, the shear modulus $\G{2}$
represents an effective secant value related to a given temperature and load
duration. For example, for data presented in~\cite{Foraboschi:2007:BFS} and the
temperature of $20^\circ$C, the value in \Sref{sec:simply_supported} corresponds
to loads with duration of $\sim 100$~s, whereas the value in
\Sref{sec:fixed_end} corresponds to loads with duration of $\sim
1,000$~s.\label{footnote}}} The problem was discretized with $\numel = 40$
elements per layer~(half if symmetry is exploited), in order to achieve the
four-digit accuracy of the mid-span deflection for $F = 200$~N, and the
tolerances of the iterative solver were set to $\epsilon_1 = \epsilon_2 =
10^{-6}$.

\begin{table}[h]
\caption{Material data for simply supported beam,
after~\cite{Asik:2005:MMB}.}
\label{tab:simply_supported_data}
\centering
\begin{tabular}{ll}
\hline
Property & Value \\
\hline
Young's modulus of glass, $\E{1} = \E{3}$ & 64.5~GPa \\
Shear modulus of glass, $\G{1} = \G{3}$ & 26.2~GPa \\
Young's modulus of PVB, $\E{2}$ & 3.61~MPa \\
Shear modulus of PVB, $\G{2}$ & 1.28~MPa \\  
\hline
\end{tabular}
\end{table}

The centerline deflections as predicted by the considered models appear in
\Fref{fig:simply_supported}(b). In this case, the deflection of the refined
formulation remains bounded by the two-layer and monolithic cases. The
response of the laminated structure is in fact closer to the monolithic beam than to the
layered approximation, thereby demonstrating that the interlayer
provides sufficient coupling to achieve the composite action. We also observe,
in agreement with assumptions of many analytical models discussed in
\Sref{sec:introduction}, the absence of non-linear effects due to
the statically-determinate character of the test.

These findings are further supported by \Tref{tab:simple_beam_mid_span}
collecting the numerical values of mid-span deflections and experimental data.
The results confirm that predictions of the numerical scheme are in a perfect
agreement with the analytical model and correlate well with the experimental
data. The limit cases, on the other hand, are too far apart to be of practical
use and exhibit errors that far exceed the difference between the refined models and
experiments.

\begin{table}[ht]
\caption{Mid-span deflections $\wc{3}$ at $\X{3} = L/2$ for the simply supported
beam subjected to $F = 50$~N.}
\label{tab:simple_beam_mid_span}
\centering
\begin{tabular}{lrrr}
\hline
Model	& Deflection~[mm]	& $\etaexp~[\%]$ & $\etaan~[\%]$ \\
\hline
Experiment$^\text{\cite{Asik:2005:MMB}}$ &
1.27 & $\times$ & -5.2 \\
Analytical model$^\text{\cite{Asik:2005:MMB}}$ & 
1.34 & 5.5 & $\times$ \\
\rev{Refined} linear & 
1.34 & 5.6 & 0.1 \\
\rev{Refined} non-linear & 
1.34 & 5.6 & 0.1 \\
Monolithic~\rev{linear} & 
0.89 & -30.1 & -33.8 \\
\rev{Two-layer linear} &
3.97 & 213 & 196 \\
\hline
\end{tabular}
\end{table}

The response of the structure to an increasing force expressed in terms of
the deflections and stresses appears in
Tables~\ref{tab:ss_deflection_comparison} and~\ref{tab:ss_stress_comparison},
respectively. The data clearly demonstrate that the response of the structure
remains linear even for larger loading; the differences appear only for
$F=200$~N and are indeed negligible. The deflections remain sufficiently
accurate with respect to both the analytical model and experiments, while stress
values display larger discrepancies. We attribute the $\sim 1.3\%$ difference
between the analytical and numerical models to the fact that A\c{s}\i{}k and
Tescan in~\cite{Asik:2005:MMB} assume glass to deform exclusively in bending and
the PVB layer in shear only, while the present approach accounts for both
effects simultaneously in all layers. This error is still significantly smaller
compared to the one measured against the experimental data, which might be
attributed to inaccuracies in the experiment as explained in detail
in~\cite{Asik:2005:MMB}.

\begin{table}[h]
\caption{Comparison of mid-span deflections $\wc{3}$ at $\X{3} = L/2$ for the
simply supported beam.}
\label{tab:ss_deflection_comparison}
\centering
\begin{tabular}{ccccccccc}
\hline
Load~[N] & 
\multicolumn{2}{c}{Reference data$^\text{\cite{Asik:2005:MMB}}$} &
\multicolumn{3}{c}{Linear model} & 
\multicolumn{3}{c}{Non-linear model} \\
& $w_\mathrm{exp}$~[mm] & $w_\mathrm{an}$~[mm] &
$w$~[mm] & $\etaexp~[\%]$ & $\etaan~[\%]$ &
$w$~[mm] & $\etaexp~[\%]$ & $\etaan~[\%]$ \\
\hline
50	& 1.27 & 1.34 & 1.34 & 5.6 & 0.1 & 1.34	& 5.6 & 0.1 \\
100	& 2.55 & 2.69 & 2.68 & 5.2 & -0.3 & 2.68 & 5.1 & -0.3 \\
150	& 4.12 & 4.03 & 4.02 & -2.3 & -0.1 & 4.02 & -2.5 & -0.3 \\
200	& 5.57 & 5.38 & 5.37 & -3.7	& -0.3 & 5.35 & -4.0 & -0.6 \\
\hline
\end{tabular}
\caption{Comparison of extreme stresses $S\lay{3}_\rev{\mathrm{bot}}$ at $\X{3}
= L/2$ for the simply supported beam.}
\label{tab:ss_stress_comparison}
\begin{tabular}{ccccccccc}
\hline
Load~[N] & 
\multicolumn{2}{c}{Reference data$^\text{\cite{Asik:2005:MMB}}$} &
\multicolumn{3}{c}{Linear model} & 
\multicolumn{3}{c}{Non-linear model} \\
& $S_\mathrm{exp}$~[MPa] & $S_\mathrm{an}$~[MPa] &
$S$~[MPa] & $\etaexp~[\%]$ & $\etaan~[\%]$ &
$S$~[MPa] & $\etaexp~[\%]$ & $\etaan~[\%]$ \\
\hline
50  & 9.55  & 7.23  & 7.14  & -25.3 & -1.3 & 7.14  & -25.2 & -1.3 \\
100	& 12.34 & 14.45 & 14.27 &  15.7	& -1.2 & 14.28 & 15.7 & -1.2 \\
150	& 21.89	& 21.68	& 21.41	& -2.2 & -1.2 & 21.42 & -2.2 & -1.2 \\
200	& 26.27	& 28.90	& 28.55 &  8.6 & -1.2 & 28.55 & 8.7	& -1.2 \\
\hline
\end{tabular}
\end{table}

\begin{figure}[h]
\begin{tabular}{ll}
\includegraphics[width=.49\textwidth]{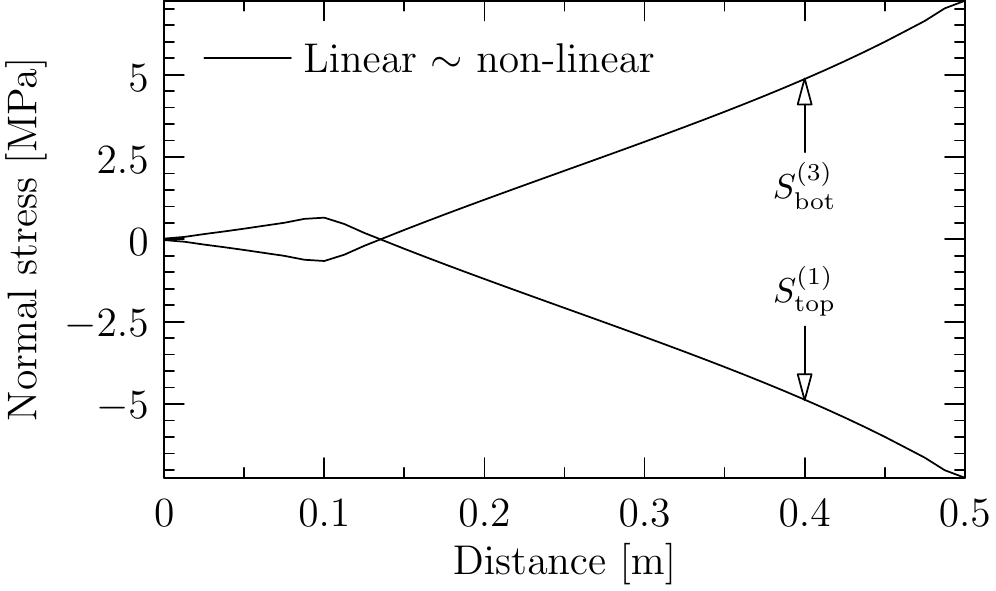} & 
\includegraphics[width=.49\textwidth]{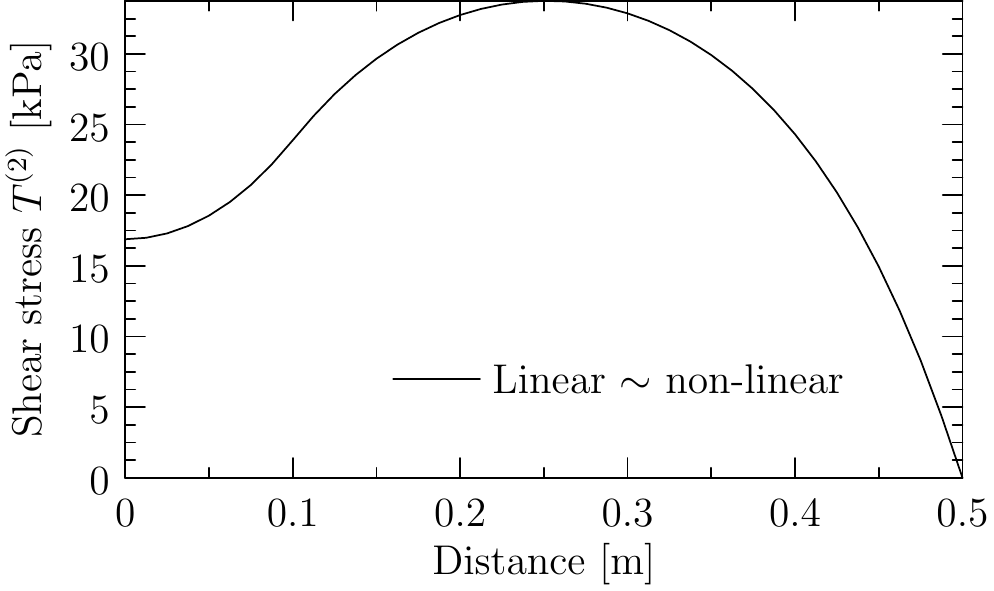} \\[-2mm]
(a) & (b)
\end{tabular}
\caption{\rev{Stress distributions in simply supported three-layer beam;
(a)~normal and (b)~shear stresses for $F=50$~N.}}
\label{fig:stress_distrib_simple}
\end{figure}

\rev{The negligible difference between the linear and non-linear models is
finally confirmed by the identical values of the normal and shear
stresses, recall~\eqref{eq:stresses}, shown in \Fref{fig:stress_distrib_simple}.
Their distribution, however, differs significantly from any single-layered
approximation, as manifested by, e.g. the presence of non-zero shear and normal
stresses \rev{at free ends} of the beams and the non-uniform distribution of
shear stresses along the beam axis. We also observe, in agreement
with~\cite{Asik:2005:MMB}, that the distribution of extreme normal stresses is
anti-symmetric in the vertical direction, i.e. $S_\mathrm{top}\lay{1} = -
S_\mathrm{bot}\lay{3}$, and that shear strains in the interlayer ($\sim 3\%$)
are significantly larger than normal strains in the glass~($\approx
10^{-4}$).}

\subsection{Fixed-end beam}\label{sec:fixed_end}

\begin{figure}[h]
\begin{tabular}{ll}
\includegraphics[width=.49\textwidth]{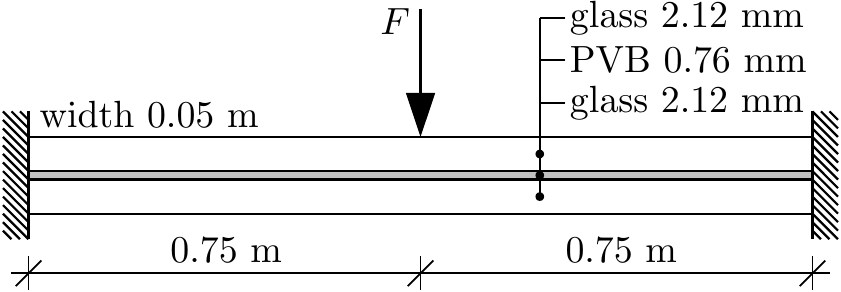} & 
\includegraphics[width=.49\textwidth]{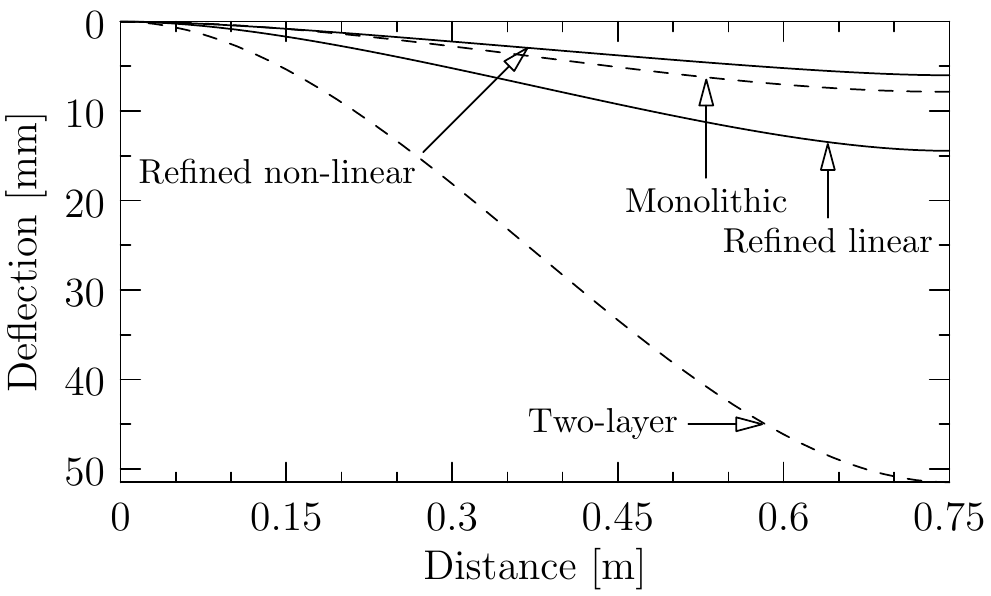} \\[-2mm]
(a) & (b)
\end{tabular}
\caption{Fixed-end three-layer beam; (a)~experiment setup and
(b)~centerline deflection for $F = 15$~N.}
\label{fig:fe_setup}
\end{figure}

Effects of geometric non-linearity are illustrated by means of a thin $1.5$~m
long three-layer beam of thicknesses 2.12--0.76--2.12 mm subjected to a
concentrated force at the mid-span with intensity ranging from $15$ to 150~N,
\rev{\Fref{fig:fe_setup}}(a). The tolerances for the Newton method were set to
the same value as in the previous example, and so were the material constants for
the glass layers. For the PVB interlayer, we used $\E{2} = 2.8$~MPa and $\G{2} =
1$~MPa.$^\text{\ref{footnote}}$ The reference data from~\cite{Asik:2005:MMB}
include the results of the analytical model, obtained by a finite-difference iterative
solver~\cite{Asik:2003:LGP}, and of detailed two-dimensional large-deformation
finite element simulations. To make a direct comparison to these values, we
employ the same number of elements per layer, $\numel = 150$, but analogous
results are obtained for coarser discretizations.

The centerline deflections for geometrically linear and non-linear refined beam
formulations appear in \rev{\Fref{fig:fe_setup}}(b) and are compared with the
response of equivalent monolithic and two-layer formulations. While the response of the
linear model still falls within the monolithic--two-layer bounds~(with a
closer proximity to the monolithic beam), the deflection of the fully non-linear model
is considerably smaller that for the monolithic case. Moreover, the gap between the
monolithic and two-layer cases becomes even more pronounced than in the
previous example.

\begin{table}[ht]
\centering
\caption{Mid-span deflections $\wc{3}$ at $\X{3} = L/2$ for the fixed beam
subjected to $F = 15$~N.}
\label{tab:fe_mid_span_deflection}
\begin{tabular}{lrrr}
\hline
Model	& Deflection~[mm]	& $\etanum~[\%]$ & $\etaan~[\%]$ \\
\hline
2D finite element model$^\text{\cite{Asik:2005:MMB}}$ &
5.92 & $\times$ & 0.0 \\
Analytical model$^\text{\cite{Asik:2005:MMB}}$ & 
5.92 & 0.0 & $\times$ \\
\rev{Refined} linear & 
14.44 & 144 & 144 \\
\rev{Refined} non-linear & 
6.00 & 1.3 & 1.3 \\
Monolithic~\rev{linear} & 
7.85 & 32.6 & 32.6 \\
\rev{Two-layer linear} &
51.48 & 770 & 770 \\
\hline
\end{tabular}
\end{table}

The detailed numerical values presented in \Tref{tab:fe_mid_span_deflection}
further support our findings. We see that the layered assumption as well as
small-strain hypothesis are too conservative and lead to 
highly inefficient designs, as their errors exceed $100\%$. The accuracy of the
monolithic approximation is comparable to the simply supported setup.

Basically the same conclusions follow from the response of the structure to an
increasing load, Tables~\ref{tab:fe_deflection_comparison}
and~\ref{tab:fe_stress_comparison}. For the largest load of $F = 150$~N, the
error of the linear laminated model reaches $\sim 800\%$ for the deflections and
$\sim 250\%$ for the extreme stresses. The finite-strain formulation remains
accurate in the whole range of loading, and the errors with respect to the
detailed two-dimensional finite element model do not exceed $\sim 1\%$ for the
deflections and $\sim 2\%$ for stresses. In fact, it slightly outperforms the
analytical model~\cite{Asik:2005:MMB} in terms of the stress values, probably
due to the more refined representation of the deformation in individual layers
as discussed above.

\begin{table}[h]
\caption{Comparison of mid-span deflections $\wc{3}$ at $\X{3} = L/2$ for the
fixed-end beam.}
\label{tab:fe_deflection_comparison}
\centering
\begin{tabular}{ccccccccc}
\hline
Load~[N] & 
\multicolumn{2}{c}{Reference data$^\text{\cite{Asik:2005:MMB}}$} &
\multicolumn{3}{c}{Linear model} & 
\multicolumn{3}{c}{Non-linear model} \\
& $w_\mathrm{an}$~[mm] & $w_\mathrm{num}$~[mm] &
$w$~[mm] & $\etaan~[\%]$ & $\etanum~[\%]$ &
$w$~[mm] & $\etaan~[\%]$ & $\etanum~[\%]$ \\
\hline
15  &  5.92 &  5.92	&  14.44 & 143.9 & 143.9 &  6.00 & 1.3 & 1.3 \\
30  &  8.10	&  8.10	&  28.88 & 256.6 & 256.6 &  8.17 & 0.8 & 0.8 \\
45  &  9.60	&  9.60	&  43.32 & 351.3 & 351.3 &  9.66 & 0.6 & 0.6 \\
60  & 10.78	& 10.78 &  57.76 & 435.9 & 435.9 & 10.83 & 0.5 & 0.5 \\
90  & 12.64	& 12.63	&  86.65 & 585.5 & 586.0 & 12.68 & 0.3 & 0.4 \\
120	& 14.10	& 14.09	& 115.53 & 719.4 & 719.9 & 14.14 & 0.3 & 0.3 \\
150	& 15.34	& 15.32	& 144.41 & 841.4 & 842.6 & 15.36 & 0.1 & 0.3 \\
\hline
\end{tabular}
\caption{Comparison of extreme stresses $S\lay{3}_\rev{\mathrm{bot}}$ at $\X{3}
= L/2$ for the fixed-end beam.}
\label{tab:fe_stress_comparison}
\begin{tabular}{ccccccccc}
\hline
Load~[N] & 
\multicolumn{2}{c}{Reference data$^\text{\cite{Asik:2005:MMB}}$} &
\multicolumn{3}{c}{\rev{Linear model}} & 
\multicolumn{3}{c}{Non-linear model} \\
& $S_\mathrm{an}$~[MPa] & $S_\mathrm{num}$~[MPa] &
$S$~[MPa] & $\etaan~[\%]$ & $\etanum~[\%]$ &
$S$~[MPa] & $\etaan~[\%]$ & $\etanum~[\%]$ \\
\hline
15  & 12.87 & 12.46 &  19.51 &  51.6 &  56.6 & 12.60 & -2.1 & 1.1 \\
30  & 20.69 & 19.89 &  39.02 &  88.6 &  96.2 & 20.12 & -2.7	& 1.2 \\
45  & 27.13	& 25.94	&  58.53 & 115.7 & 125.6 & 26.28 & -3.2	& 1.3 \\
60  & 32.82	& 31.25	&  78.03 & 137.8 & 149.7 & 31.69 & -3.4	& 1.4 \\
90  & 42.82	& 40.51	& 117.05 & 173.4 & 188.9 & 41.18 & -3.8	& 1.6 \\
120	& 51.68	& 48.64	& 156.07 & 202.0 & 220.9 & 49.53 & -4.2	& 1.8 \\
150	& 59.76	& 56.00	& 195.09 & 226.4 & 248.4 & 57.13 & -4.4	& 2.0 \\
\hline
\end{tabular}
\end{table}

\begin{figure}[h]
\begin{tabular}{ll}
\includegraphics[width=.49\textwidth]{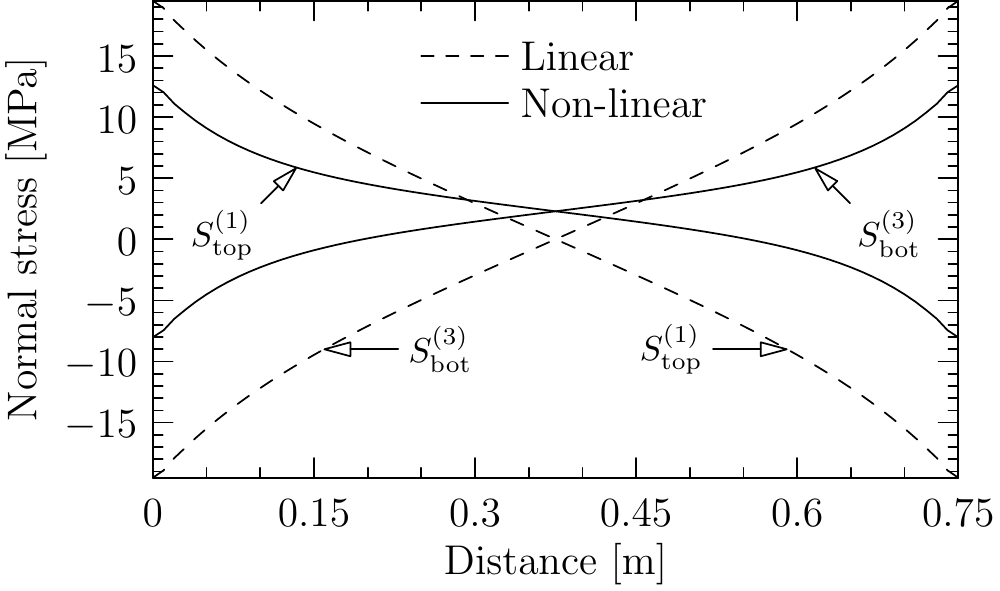} & 
\includegraphics[width=.49\textwidth]{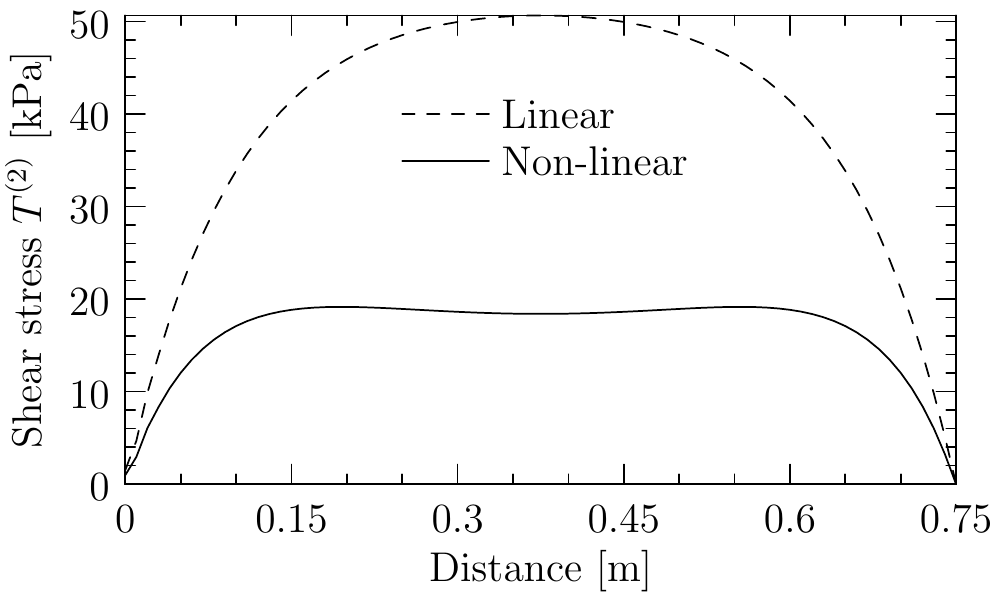} \\[-2mm]
(a) & (b)
\end{tabular}
\caption{\rev{Stress distributions in fixed-end three-layer beam; (a)~normal and
(b)~shear stresses for $F=15$~N.}}
\label{fig:stress_distrib}
\end{figure}

\rev{Unlike the previous example, the stress distributions for the linear and
non-linear models differ to a significant amount. As for the extreme normal
stresses, \Fref{fig:stress_distrib}(a), in the non-linear model their magnitude
reaches $\sim 65\%$ of the value obtained by the linear one and their
distribution is no longer antisymmetric in the thickness direction, i.e.
$S_\mathrm{top}\lay{1} \neq - S_\mathrm{bot}\lay{3}$. These effects are due to
additional membrane stresses acting in the glass layers. The same mechanism
reduces the magnitude of the shear stresses in the interlayer to $\sim 40\%$ of
value determined for the linear analysis, \Fref{fig:stress_distrib}(b). We note
again that these observations are in an agreement with the results reported by
A\c{s}\i{}k and Tescan~\cite{Asik:2005:MMB}.}

As the final check of our implementation, in \Tref{tab:convergence_progress} we
present the convergence progress for the load level $F = 150$~N. In order to
investigate reliability of the method, we expose the structure to the full load,
instead of applying it incrementally as in~\cite{Asik:2005:MMB}. The results
confirm significant degree of non-linearity in the structural response. After
the first iteration, corresponding to the linear model, the structure is in an
out-of-equilibrium state and the layer compatibility is violated. In the
following iterations, the residuals are gradually reduced \rev{in a
non-monotonic way} until the ninth iteration, after which the method exhibits
quadratic convergence in the equilibrium residual $\eta_1$ and super-linear
convergence in the compatibility residual $\eta_2$. This behavior is in full
agreement with available results for the Newton method,
e.g.~\cite[Theorem~13.6]{Bonnans:2003:NOTPA}, and also explains the convergence
difficulties of the heuristic finite-difference solver reported
in~\cite{Asik:2005:MMB}.

\begin{table}[h]
\caption{Convergence of residuals, \rev{defined by~\Eref{eq:residuals}}, for
$F = 150$~N.}
\label{tab:convergence_progress}
\centering
\begin{tabular}{crr}
\hline 
Iteration $k$ & 
\multicolumn{1}{c}{$\ite{k}\eta_1$} & 
\multicolumn{1}{c}{$\ite{k}\eta_2$} \\
\hline 
 1 & $8.49\times 10^2$ & $7.94\times 10^{-1}$ \\
 2 & $1.50\times 10^3$ & $4.65\times 10^{-1}$ \\
 3 & $1.02\times 10^2$ & $6.12\times 10^{-2}$ \\
 4 & $2.07\times 10^2$ & $5.61\times 10^{-2}$ \\
 5 & $2.31\times 10^1$ & $1.11\times 10^{-2}$ \\
 6 & $2.43\times 10^1$ & $7.53\times 10^{-3}$ \\
 7 & $4.93\times 10^0$ & $2.58\times 10^{-3}$ \\
 8 & $1.41\times 10^0$ & $8.17\times 10^{-4}$ \\
 9 & $1.38\times 10^{-1}$ & $8.23\times 10^{-5}$ \\
10 & $1.58\times 10^{-3}$ & $8.33\times 10^{-7}$ \\
11 & $2.53\times 10^{-7}$ & $1.18\times 10^{-10}$ \\
\hline
\end{tabular}
\end{table}

\section{Conclusions}\label{sec:conclusions}
%%%%%%%%%%%%%%%%%%%%%
In this paper, we have presented an efficient approach to the analysis of the
laminated glass beams. Based on the Mau theory for layered structures, it treats
each layer separately and enforces the inter-layer compatibility by the Lagrange
multipliers. In our implementation, we utilize a reliable finite element
formulation of the Reissner finite-strain beam theory due to Ibrahimbegovi\'{c}
and Frey to discretize individual layers, and solve the resulting system of
equations iteratively by the Newton method with consistent linearization. On the
basis of the performed simulations, we conjecture that

\begin{itemize}
  \item in the absence of membrane effects, the formulation reduces exactly to
  the small-strain model introduced in our previous work, 
  \item although the discretization is based on the lowest-order polynomial
  basis functions, the method provides results with accuracy comparable to
  the detailed two-dimensional large-strain finite element simulations, 
  \item the Newton method exhibits a reliable super-linear convergence even for
  high degrees of non-linearity. 
\end{itemize}

Extension of the current framework to include temperature- and time-dependent
properties of the interlayer will be reported independently.

\paragraph{Acknowledgments}
The authors thank Professor Ji\v{r}\'{\i} \v{S}ejnoha of CTU in Prague
for his helpfully comments on the original manuscript. This work was supported by
the Czech Science Foundation, projects No.~P105/12/0331~(AZ) and
No.~P105/11/0224~(M\v{S}), and by the Grant Agency of the Czech Technical
University in Prague, project No.~SGS13/034/OHK1/1T/11~(JZ, M\v{S}). In
addition, JZ acknowledges the partial support of the European Regional
Development Fund under the IT4Innovations Centre of Excellence, project
No.~CZ.1.05/1.1.00/02.0070.

\providecommand{\bysame}{\leavevmode\hbox to3em{\hrulefill}\thinspace}

\appendix

\section{Sensitivity analysis}\label{app:sensitivity_analysis}

The expression for the internal nodal forces follows directly from
\Eref{eq:internal_forces}$_1$. After certain manipulations
we arrive at, cf.~\cite{Kucerova:2003:DEA}
\begin{equation}
\Mfinte{\el}\lay{i}
=
\begin{bmatrix}
\finte{e,1}\lay{i} \\
\finte{e,2}\lay{i} \\
\finte{e,3}\lay{i} \\
\finte{e,4}\lay{i} \\
\finte{e,5}\lay{i} \\
\finte{e,6}\lay{i} 
\end{bmatrix}
=
\left[
\begin{array}{c}
- \E{i} \A{i} \Sc{i}_\el \cos\beta\lay{i}_\el 
- \G{i} \As{i} \Gc{i}_\el \sin\beta\lay{i}_\el  \\
  \E{i} \A{i} \Sc{i}_\el \sin\beta\lay{i}_\el 
- \G{i} \As{i} \Gc{i}_\el \cos\beta\lay{i}_\el \\
- \half (\eL{\el}{i} + \Duc{i}{\el})
  \finte{e,2}\lay{i} 
+ \half \Dwc{i}{\el} \finte{e,1}\lay{i} 
- \E{i} \I{i} \Kc{i}_\el
\\ 
- \finte{e,1}\lay{i} \\
- \finte{e,2}\lay{i} \\
- \half (\eL{\el}{i} + \Duc{i}{\el} )
\finte{e,2}\lay{i} 
+ \half \Dwc{i}{\el}\finte{e,1}\lay{i} 
+ 
\E{i} \I{i} \Kc{i}_\el 
\end{array}
\right].
\label{eq:f_int_e}
\end{equation}
By an analogous procedure one obtains from \Eref{eq:internal_forces}$_2$ 
\begin{equation}
\M{K}_{\mathrm t, e}\lay{i} =
\left[
\begin{array}{rrrrrr}
 \Kt{e,11}\lay{i} &  \Kt{e,12}\lay{i} &  \Kt{e,13}\lay{i} & - \Kt{e,11}\lay{i} & - \Kt{e,12}\lay{i} &  \Kt{e,13}\lay{i} \\
 \Kt{e,12}\lay{i} &  \Kt{e,22}\lay{i} &  \Kt{e,23}\lay{i} & - \Kt{e,12}\lay{i} & - \Kt{e,22}\lay{i} &  \Kt{e,23}\lay{i} \\
 \Kt{e,13}\lay{i} &  \Kt{e,23}\lay{i} &  \Kt{e,33}\lay{i} & - \Kt{e,13}\lay{i} & - \Kt{e,23}\lay{i} &  \Kt{e,36}\lay{i} \\
-\Kt{e,11}\lay{i} & -\Kt{e,12}\lay{i} & -\Kt{e,13}\lay{i} &   \Kt{e,11}\lay{i} &   \Kt{e,12}\lay{i} & -\Kt{e,13}\lay{i} \\
-\Kt{e,12}\lay{i} & -\Kt{e,22}\lay{i} & -\Kt{e,23}\lay{i} &   \Kt{e,12}\lay{i} &   \Kt{e,22}\lay{i} & -\Kt{e,23}\lay{i} \\
 \Kt{e,13}\lay{i} &  \Kt{e,23}\lay{i} &  \Kt{e,36}\lay{i} & - \Kt{e,13}\lay{i} & - \Kt{e,23}\lay{i} &  \Kt{e,33}\lay{i} 
\end{array}
\right],
\label{eq:K_e}
\end{equation}
where the individual entries read
\begin{align*}
\Kt{e,11}\lay{i} 
& =  
\ppd{\finte{e,1}\lay{i}}{\uce{i}{\el,1}} 
= 
\frac{1}{\eL{e}{i}} 
\left( 
  \E{i} \A{i} \cos^2 \beta_\el\lay{i} 
  + 
  \G{i} \As{i} \sin^2 \beta_\el\lay{i} 
\right), 
\\ 
\Kt{e,12}\lay{i} 
& = 
\ppd{\finte{e,1}\lay{i}}{\wce{i}{\el,1}} 
= 
\frac{1}{2\eL{e}{i}} 
\left( 
  -\E{i} \A{i} 
  + 
  \G{i} \As{i} 
\right) 
\sin 2\beta_\el\lay{i}, 
\\
\Kt{e,13}\lay{i} 
& = 
\ppd{\finte{e,1}\lay{i}}{\rote{i}{\el,1}}
= 
\half 
\left[ \left( 
  \E{i}\A{i} 
  - 
  \G{i}\As{i} 
\right) 
\left( 
  \Sc{i}_\el \sin \beta_\el\lay{i} 
  + 
  \Gc{i}_\el \cos \beta_\el\lay{i}
\right) 
- 
\G{i} \As{i} \sin \beta_\el\lay{i} 
\right],
\\
\Kt{e,22}\lay{i} 
& =  
\ppd{\finte{e,2}\lay{i}}{\wce{i}{\el,1}}
= 
\frac{1}{\eL{e}{i}} 
\left( 
  \E{i} \A{i} \sin^2 \beta_\el\lay{i} 
  + 
  \G{i} \As{i} \cos^2 \beta_\el\lay{i} 
\right), 
\\
\Kt{e,23}\lay{i} 
& = 
\ppd{\finte{e,2}\lay{i}}{\rote{i}{\el,1}} 
= 
\half 
\left[ \left( 
  \E{i} \A{i} 
  - 
  \G{i} \As{i} 
\right) 
\left( 
  \Sc{i}_\el \cos \beta_\el\lay{i} 
  - 
  \Gc{i}_\el \sin \beta_\el\lay{i}
\right) 
- 
\G{i} \As{i} \cos \beta_\el\lay{i} 
\right],
\\
\Kt{e,33}\lay{i} 
& = 
\ppd{\finte{e,3}\lay{i}}{\rote{i}{e,1}} 
= 
\half 
\left[ -\left( 
  \eL{e}{i} 
  + 
  \Duc{i}{\el}
\right) \Kt{e,23}\lay{i} 
+ 
\Dwc{i}{\el}
\Kt{e,13}\lay{i} \right] 
+ 
\frac{\E{i}\I{i}}{\eL{e}{i}},
\\
\Kt{e,36}\lay{i} 
&= 
\ppd{\finte{e,3}\lay{i}}{\rote{i}{e,2}} 
= 
\Kt{e,33}\lay{i} 
- 
\frac{2 \E{i} \I{i}}{\eL{e}{i}}.
\end{align*}

The remaining terms in the system~\eqref{eq:system} originate from the
compatibility conditions~\eqref{eq:CC2}. In
particular, the matrix $\M{C}$ is analogous to the small-strain tying
condition~\cite[Section~4]{Zemanova:2008:SNM}. The block of $\M{C}$,
associated with a node $j$ and layers $i$ and $(i+1)$ attains the form
\begin{eqnarray}
\M{C}\lay{i,i+1}_{j} 
= 
\begin{bmatrix}
  1 & 0 & \half \h{i}\cos \rot{i}_j & \cdots & 
 -1 & 0 & \half \h{i+1} \cos \rot{i+1}_j \\ 
  0 & 1 &-\half \h{i} \sin \rot{i}_j & \cdots &  
  0 &-1 &-\half \h{i+1} \sin \rot{i+1}_j
\end{bmatrix}.
\end{eqnarray}
The second derivatives of the compatibility conditions quantify their
contributions to the tangent stiffness
\begin{eqnarray}
\M{K}_{\lambda, j} \lay{i,i+1} 
=
\ppd{^2 c_X\lay{i,i+1}}{{\Md\lay{i,i+1}_{j}}^2} \lambda_{X,j}\lay{i,i+1} + 
\ppd{^2 c_Z\lay{i,i+1}}{{\Md\lay{i,i+1}_{j}}^2} \lambda_{Z,j}\lay{i,i+1}.
\end{eqnarray}
This additional term is expressed as
\begin{eqnarray}
\M{K}_{\lambda, j} \lay{i,i+1} = 
\begin{bmatrix}
  0 & 0 & 0 & \cdots & 0 & 0 & 0 \\
  0 & 0 & 0 & \cdots & 0 & 0 & 0 \\
  0 & 0 & 
  K_{\lambda, \rot{i}_j} \lay{i,i+1}
            & \cdots & 0 & 0 & 0 \\
  \vdots & \vdots & \vdots & \ddots & \vdots & \vdots & \vdots \\
  0 & 0 & 0 & \cdots & 0 & 0 & 0 \\
  0 & 0 & 0 & \cdots & 0 & 0 & 0 \\
  0 & 0 & 0 & \cdots & 0 & 0 & 
  K_{\lambda, \rot{i+1}_j} \lay{i,i+1}
\end{bmatrix},
\end{eqnarray}
with non-zero entries provided by
\begin{subequations}
\begin{align}
K_{\lambda, \rot{i}_j}\lay{i,i+1} 
& = 
-\half \h{i} ( \sin\rot{i}_j \lambda_{X,j}\lay{i,i+1}
+ \cos\rot{i}_j \lambda_{Z,j}\lay{i,i+1} ),
\\
K_{\lambda, \rot{i+1}_j}\lay{i,i+1} 
& = 
-\half \h{i+1} (  \sin\rot{i+1}_j \lambda_{X,j}\lay{i,i+1}
+ \cos\rot{i+1}_j \lambda_{Z,j}\lay{i,i+1} ).
\end{align}
\end{subequations}

%\printnomenclature

\end{document}